# Tailoring the Cyber Security Framework:
# How to Overcome the Complexities of Secure Live Virtual Machine Migration in Cloud Computing

Hanif Deylami, Jairo Gutierrez, Roopak Sinha
*Auckland University of Technology*
*New Zealand*
hmohadde@aut.ac.nz, jairo.gutierrez@aut.ac.nz, roopak.sinha@aut.ac.nz

## Abstract

*This paper proposes a novel secure live virtual machine migration framework by using a virtual trusted platform module instance to improve the integrity of the migration process from one virtual machine to another on the same platform. The proposed framework, called Kororā, is designed and developed on a public infrastructure-as-a-service cloud-computing environment and runs concurrently on the same hardware components (Input/Output, Central Processing Unit, Memory) and the same hypervisor (Xen); however, a combination of parameters needs to be evaluated before implementing Kororā. The implementation of Kororā is not practically feasible in traditional distributed computing environments. It requires fixed resources with high-performance capabilities, connected through a high- speed, reliable network. The following research objectives were determined to identify the integrity features of live virtual machine migration in the cloud system:*

- *To understand the security issues associated with cloud computing, virtual trusted platform modules, virtualization, live virtual machine migration, and hypervisors;*
- *To identify the requirements for the proposed framework, including those related to live VM migration among different hypervisors;*
- *To design and validate the model, processes, and architectural features of the proposed framework;*
- *To propose and implement an end-to-end security architectural blueprint for cloud environments, providing an integrated view of protection mechanisms, and then to validate the proposed framework to improve the integrity of live VM migration.*

*This is followed by a comprehensive review of the evaluation system architecture and the proposed framework state machine. The overarching aim of this paper, therefore, is to present a detailed analysis of the cloud computing security problem, from the perspective of cloud architectures and the cloud service delivery models. Based on this analysis, this study derives a detailed specification of the cloud live virtual machine migration integrity problem and key features that should be covered by the proposed framework.*

## 1. INTRODUCTION

The word "Cloud" is a metaphor describing the web as space where computing has been preinstalled and exists as a service. Many companies, both large and small, are contemplating a migration to cloud computing (CC) to leverage the significant potential of this new paradigm [1-3]. Government agencies, small and medium-sized enterprises, and large organizations already make significant use of CC and they are spending considerable amounts of money, resources, and time on delivering secure services using information and communication technologies [4]. Security is crucial and it is one of the main challenges for CC adoption, as many surveys show [5]. Systems become significantly more susceptible to several cyber attacks when they move to cloud platforms, especially when this move is based on a lack of adoption of cloud-native models and the required adjustment in organizational processes to align with the features and capabilities of the chosen cloud platforms [2].

Virtualization is a technology that provides the ability to automate and orchestrate multiple, tightly isolated IT processes related to on-demand provisioning on a single piece of server hardware to create a virtual computer system or "Virtual Machine" (VM). With respect to virtualization technologies, a physical server can be divided into several isolated execution environments by developing a layer (i.e., VM monitor or hypervisor) on top of the hardware resources or operating systems (OSs); thus, a physical database can be divided into several separate execution environments with the help of virtualization technologies. The server's execution environments (i.e., VMs) run in parallel without interruption. A VM, also called a guest machine, is a virtual representation, or software emulation of a hardware platform that provides a virtual operating environment for guest OSs. The task of moving a VM from one physical hardware environment to another is called migration. If the migration is carried out in such a way that the connected clients perceive no service interruption, it is considered a "live" migration. For example, database consolidation is made easier if VMs do not have to be shut down before they are transferred. The method is also used for administrative purposes; for instance, if a server needs to be taken off-line for some reason, live transferring of VMs to other hosts can be used to pass running VMs between cloud sites over wide-area communication networks.

A VM migration can occur in two ways: live and offline. In a live VM migration, the VMs are transferred from a source host to a destination host while they are running. After a successful VM migration, the source host removes the memory pages of the migrated VM. During a live VM job migration, there is considerable potential for compromise through malicious activities while information such as memory pages is copied from the host and transferred to the destination, presenting security risk(s) with regard to data integrity and confidentiality. The owner of the VM must have a framework to ensure live VM migration data integrity at both ends of the migration process. In other words, there is a need to ensure a clear memory portion is assigned to an incoming VM on the destination host, separated from previous VM data or malicious codes, and to secure the removal of the memory data of the outgoing VM. This situation might make organizations and businesses reluctant to switch to using the cloud because of the potential attacks on their assets.

This paper investigates the possibility of misuse of migrating VM's data either in transit or present at source and destination during the live VM migration process. It then proposes a novel framework for a secure live VM migration by using a Virtual Trust Platform Model (vTPM) agent and four other agents: input/output, data plane, integrity analyzer, and data organization. While existing studies [6-9] have established a live VM migration framework for cloud systems integrity, an examination of the different types of research has identified a lack of empirical evidence and knowledge regarding which issues are the most important for these areas. In this paper, the relative significance of the identified issues is determined first, to address the two related research questions listed below, and then the importance of the identified issues is discussed in the rest of the paper. *Research Question 1*: What are the opportunities and challenges for live VM migration in CC, with respect to the essential system attributes and essential system characteristics?, and *Research Question 2*: What are the characteristics of the proposed framework that is on the public cloud instead of in a traditional on-premises data center? According to Kitchenham et al. [10], a systematic literature review method is one of the best ways to identify and prioritize issues for decision making and to sort large volumes of references. This method assists in identifying the research questions and issues associated with the research topic. The overarching aim of this paper is to develop and design a secure live VM migration framework to help cloud service providers (CSPs) improve integrity protection in live VM migration from one VM to another in the same platform (with the same hardware features and the same hypervisor [Xen hypervisor]).

The remainder of this paper is structured as follows. Section 2 discusses the related work and motivation for this research. Section 3 explains the design of the framework system architecture and its agents. Section 4 presents the evaluation system architecture: that is, the state machine. Finally, Section 5 summarizes the paper and discusses future work.

## 2. RELATED WORK AND MOTIVATION

Critical concerns for cloud users involve protecting workloads and data in the cloud and from the cloud, and ensuring trust and integrity for VM images launched on a service provider's cloud [11]. For live VM and workload data protection, cloud-user organizations need a framework to securely place and use their workloads and data in the cloud. Current provisioning and deployment frameworks include either storing the VM and application images and data in the clear (i.e., unencrypted) or having these images and data encrypted using keys controlled by the service provider, which are likely applied uniformly to all the tenants.

Live VM migration [12] in the inter-cloud is a new way of looking at VM migration. It allows the migration of VMs not only between data centers of the same cloud but also between servers on different clouds. The driving force behind live VM migration between clouds is to decrease the workload on a particular cloud and reduce the congestion of its network. The key point of a planned migration is to take snapshots that preserve the state and data of a VM at any given time. With these snapshots of a VM, an image of the VM in each state is copied and stored. The snapshot is then migrated to the destination cloud, where the hypervisor creates a new VM with the same configuration as the snapshot. The source cloud redirects the incoming traffic of its VM to the destination VM soon after the target VM is up and running.

Data deduplication [6] is a live VM migration technique that prevents large chunks of data from migrating, thereby reducing migration time. This operates on the concept of only selected memory material that has been altered on the source server being transferred. Thus, the phase of migration involves only those parts of the VM that were updated at the source end. A Dirty Block Tracking (DBT) mechanism and a new diff format are the two major components of data deduplication. The role of DBT is to record all the operations that cause changes in the picture of the VM disk, while the diff format is used to store the reported data.

DBT monitors and labels each changed disk page as a dirty file. Only the pages identified by the DBT are migrated to the storage; the rest is left behind. Data deduplication is beneficial for VMs undergoing multiple migrations, resulting in multiple destination servers. As it reduces the migration time by a factor of 10, it is one of the most effective techniques for live VM migration.

Yang et al. [7] suggest an Input/Output (I/O) Outsourcing scheme for Workload-Aware, (WAIO) to improve the efficiency of live processing for VM migration. During the migration, WAIO effectively outsources the working set of the VM to a surrogate device and creates a separate I/O path to serve VM I/O requests. The VM live storage migration process can be performed on the original storage by outsourcing VM I/O requests from the original storage to the surrogate device, without interfering with them, while the outsourced VM I/O requests are serviced separately and thus, much faster. This lightweight WAIO prototype implementation and extensive trace-driven experiments show that WAIO significantly improves the I/O performance of the VM during the migration process compared with the existing DBT migration approach. In addition, WAIO allows the hypervisor to migrate a VM at a higher speed of migration without sacrificing the I/O performance of the VM.

Riteau et al. [8] propose a live VM migration system, called Shrinker, which allows VM clusters to migrate between data centers linked via a network. Through integrating data duplication and cryptography hash functions, Shrinker reduces the data to be migrated. This operates on the principles of handling distributed information, and of allowing chunks of VMs to be migrated in multiple data centers across different servers. Shrinker is different from traditional live VM migration methods as it allows source and destination server hypervisors to interact with each other during migration.

Work on opportunistic replay [13] aims to reduce the amount of data in low bandwidth environments that are migrated. This approach keeps a record of all types of user events that occur during the execution of the VM. This information is then transferred to an identical manufactured VM and put into effect to produce almost the same state as the VM source.

Zheng et al. [9] present a novel scheduling algorithm for storage migration that can significantly improve the performance of I/O storage during wide-area migration. This algorithm is unique in that it considers the storage I/O workload of individual VMs, such as temporal location, spatial location, and popularity characteristics, to calculate efficient schedule data transfers.

Berger et al. [17] discuss a vTPM that provides trusted computing for multiple VMs running on a single platform. The key to this process is finding a way to store vTPM data encrypted in the source platform and restoring them safely in the in-destination platform, as well as a way to protect the integrity of the transferred data in the process of live vTPM-VM migration, where it is vulnerable to all the threats of data exchange over a public network. These include leakage, falsification, and loss of sensitive information contained in the VM and vTPM instances.

This paper proposes a better alternative live VM migration framework, which assigns valid but conspicuous values in the new system as "flags" for problem data. This means that when users find a flag in a certain record, they know that the migrated record contains information that could not be loaded immediately. The original data from the legacy system persist in a standard format and are connected to the new record for each such example. The user can quickly check the original source to interpret the data in a meaningful manner.

In addition, the proposed framework collects the target VM working set data over the migration period to the Kororā platform. This helps the framework process to access the data set during migration, while the I/O migration process is accessing the original disk most of the time. Consequently, it is possible to significantly reduce the traffic between I/O processes and the Kororā platform, and the overall integrity of the live VM migration can be improved.

## 3. SYSTEM ARCHITECTURE

The use of the IT security framework is supported by tools that enable service providers to bridge the gap between control requirements, technical issues, and business risks. Kororā is capable of measuring and preserving the integrity of live VMs migration in the cloud system. The expected benefits of using this framework include increasing the level of integrity among different physical hosts. Kororā allows users to check malicious files against three different malware providers' engines and it can check indicators of comparison details of hashes, URLs, IP addresses, and domains from different resources.

This section aims to explain the system requirements (representing the problem from a design point of view) through an intermediate model of logical architecture, to allocate the elements of the logical architecture model to the system elements of the Kororā physical architecture models. The proposed framework system requirements and the exact approach taken in the synthesis of solutions often depends on whether the system is an evolution of an already-understood product. The Kororā system architecture aims to meet the following system elements and system architecture requirements:

- *System Element 1 – Integrity of configuration files:* In this case, the VM image structure is such that it can represent a complete file system for a given platform integrity: for example, 'vbox' files in virtual box or '.vmx' files in VMware. Both these files can be edited by a third party to make changes in the configuration of VMs.

- *System Element 2 – Virtual hard disk integrity:* The life cycle of the VM image consists of different states. For instance, a VM image can be created, started, suspended, stopped, migrated, or destroyed. Essentially, VM images are loaded from a storage location such as a hard disk drive and run directly from a VM manager with a low level of integrity: for example, '.vmdk', '.vdi', '.ova' files. A third party can make changes to these files after running them in their own environment since it is the actual OS holding file; it would be easy to place a Trojan or malicious codes inside the files.

- *System Element 3 – The integrity of the data files on the VM, including all confidential files, and the integrity of the system files:* The VM is loaded from the storage location and the VM image may not comply with the intended settings and configurations needed for proper implementation in each environment. The VM image itself could be distorted (perhaps by an insider) or even maliciously modified. This work proposes two ways to analyze these files – "*supply the data files*" and "*system files hashsum*" – on the framework before migration and checking of the files after migration.

### 3.1. System Architecture Requirements

To apply the system design agents in the Kororā framework, the following requirements must be considered in the Xen hypervisor environment:

- 64-bit x86 computer with at least 1 GB of RAM (this can be a server, desktop, or laptop) and trusted platform module chipset on the motherboard. The TPM hardware must be activated through the BIOS.

- Intel virtualization technology or AMD-V support (optional for paravirtualization [PV], required for hardware VM and some PV optimisation).

- Sufficient storage space for the Kororā framework dom0 installation.

- Extensible firmware interface – this helps the hardware layer to select the OS and get clear of the boot loader. In addition, it helps the CSP to protect the created drivers from a reverse-engineering (back-engineering) attack.

- Software requirement cmake – this is the main additional product necessary for compiling a vTPM. To manage domains with vTPM, libxl should be used rather than 'xm' which does not support vTPM.

- Linux host (Ubuntu 12.4) must be installed on the machine.

The Kororā system architecture focuses on a hypervisor that preserves metadata using cryptography and hashing algorithms. The protected live VM migration framework based on this hypervisor was designed to identify the different attacks possible and perform an independent secure migration process. The approaches of live VM migration are generally divided into three different classes: 1) *Migration of the process*; 2) *Migration of memory*; 3) *Suspend/resume migration*. In this research, the process of live VM migration means the process of migrating a VM from a source host to a destination host without suffering any attacks. These requirements must be incorporated into the process of the secure live VM migration platform.

Before the migration starts, it is important to ensure that source hosts and destination hosts and VMs meet the requirements for migration that Kororā is trying to match and to verify whether the target is correct, and to create a cryptography rule. Effective access control policies must be provided to protect the process of live VM migration. If an unauthorized user/role begins the live VM process and initiates the migration, the use of access control lists in the hypervisor will avoid the occurrence of unauthorized activities (authorization). Using route hijacking or Address Resolution Protocol (ARP) poisoning techniques in the migration process, an attacker may initiate Man-in-the-Middle (MiTM) attacks. During live VM migration, the source and destination platforms need to perform mutual authentication in order to avoid MiTM attacks (authentication). An encrypted network must be set up so that no data can be accessed from the VM content by an intruder and any software alteration can be detected properly. This will help to prevent active attacks on live migration, such as memory manipulation, and passive attacks, such as sensitive information leakage (confidentiality and integrity). An intruder may intercept traffic and later replay it for authentication in the process of the VM migration. Therefore, the method of live VM migration should be immune to replay. For example, nonces in java applications help with the password for the migration authorization, as well as the public key of the machine where the user is sitting at, to provide the correct command that is transmitted to the server in migration to prevent playback attack (reply resistance). The source host cannot deny the VM migration activity. Using public key certificates can achieve this feature (source non-repudiation).

This framework is orthogonal to existing live migration approaches – including the Zehang et al. [9] and Mashtizadeh et al. [15] live migration patents, and the Fan Peiru [16] vTPM-VM live migration protocol – and it is a secure boost layer for most, if not all, VM live migration schemes. In addition, this framework can be used to improve the security of other VM tasks, such as those associated with the virtualization and the virtual networking layers, which may experience the same problem of data integrity as VM live storage migration. This research framework, as well as the three frameworks named above, exploit the secure live migration characteristics, but they improve the VM migration security in different ways. For example, the scheme of Zheng et al. [9] aims to significantly reduce the total amount of data transferred by exploiting the workload of the VM's locality. Rarely updated data blocks are differentiated from frequently updated data blocks in virtual disk images by analyzing the workload position. The rarely updated data blocks are transferred in the migration before the frequently updated data blocks, so that the re-transmissions of data blocks are minimized, thus reducing the total amount of data transmissions. While this current research framework secures the live VM migration, its methodology is completely different from that of Zehang [9].

Five agents of the design framework system architecture must be clarified. The responsibilities of these agents are as follows:

- *Virtual Trust Platform Model Agent:* The vTPM agent provides trusted computing for multiple VMs migration on a single platform [17]. With multiple VMs operating on a single platform, vTPM offers trusted computing. It is important to move the vTPM instance data along with its corresponding VM data to keep the VM security status synched before and after the live vTPM-VM migration process. Current live VM migration schemes only check the hosts' reliability and integrity. This poses a huge security risk for vTPM-VM migration. To solve this problem, the proposed framework uses vTPM to secure boot VM(s) over the Xen hypervisor (see Figure 1, Label 1).

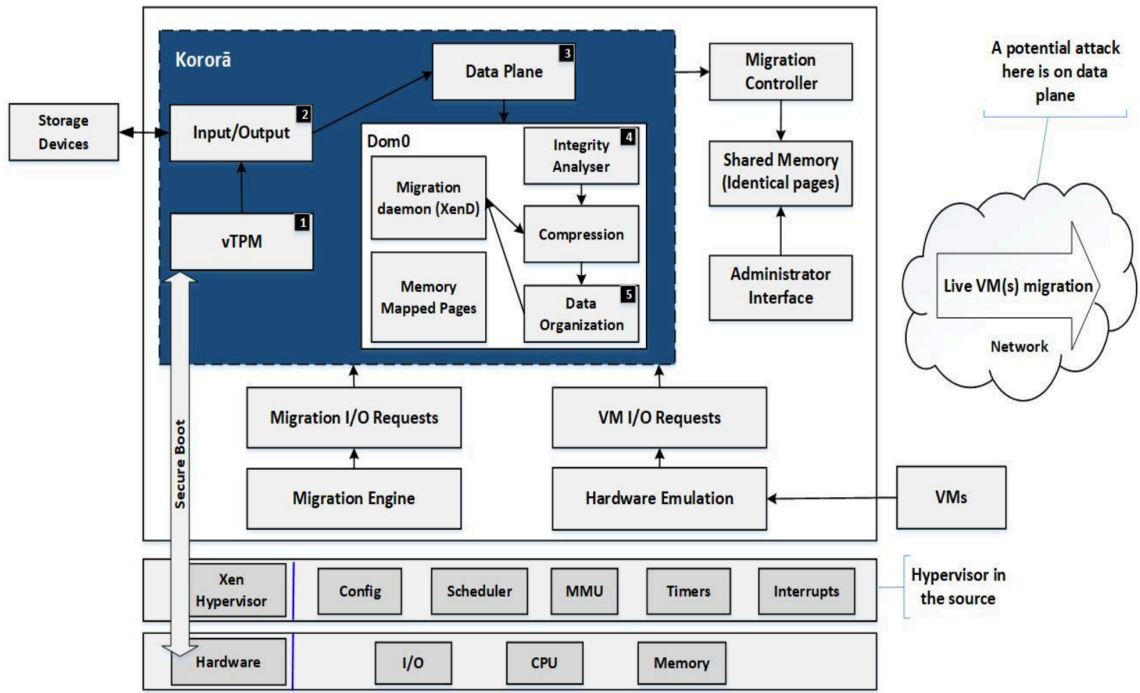

Figure 1. System Design Architecture of the Kororā

- *I/O Agent:* The I/O agent redirects the necessary I/O requests to the replacement device from the operating VM itself. To minimize I/O traffic to the original replacement device, it redirects all write requests on the replacement device [18]. Meanwhile, the I/O redirects all the popular read requests identified by the Data Plane module to the replacement device. If the replacement device has only partial data for a request, the I/O issues read requests to the original replacement device and merge the data from the original device into the replacement device. Either the original storage device [18] or the replacement device can be redirected to the read requests from the migration module. While the original storage device generates most of the virtual disk images, the replacement device provides the modified chunks (units of information that contain either control information or user data) of data. Because of the VM workload locality, most of the requests will be routed to the original storage device (see Figure 1, Label 2).

- *Data Plane Agent:* Different memory contents are moved from one host to another host in this module (e.g., kernel states and application data). The transmission channel must, therefore, be secured and protected from any attack. All migrated data are transferred as clear data without encryption in the live VM migration protocol. An attacker may, therefore, use one of the following techniques to position himself in the transmission channel to execute a MiTM attack: ARP spoofing, DNS poisoning, or route hijacking [19, 20]. These attacks are not theoretical. Tools such as Xensploit work against Xen and VMware migration [21] (see Figure 1, Label 3).

- *Integrity Analyzer Agent:* Protection of information systems is concerned with three key information properties: availability, integrity, and confidentiality. These three critical characteristics of information are major concerns throughout the commercial and military sectors. Traditionally, confidentiality has received the most attention, probably because of its importance in the military. Unlike the military security systems, the main concern of commercial security is to ensure the integrity of data is protected from unauthorized users. Availability and confidentiality are equally significant within the commercial environment, where a secure working environment is required; however, Clark and Wilson (CW) [22] propose a security model that focuses on integrity in recognized mathematical terms by a set of constraints, or a valid state when it satisfies these. Since much of the attention in the security arena has been devoted to developing sophisticated models (e.g., Bell-LaPadula model [23, 24]) and mechanisms for confidentiality, capabilities to provide confidentiality in information systems are considerably more advanced than those providing integrity.

- The integrity analyzer agent uses CW as a basic theory for specifying and analyzing an integrity policy for Kororā. Moreover, it adopts the CW model to live VM migration focusing on the subjects, objects (see Section 4), and their data exchange of users' applications to enhance the security of the live VM migration mechanism, as well as providing user convenience (see Figure 1, Label 4).

- *Data Organization Agent:* In the virtual disk images, the data organization agent monitors the popularity of reading requests from the live VM itself. Only the popular data blocks that will be read are outsourced to the replacement device. Since the replacement device serves all write requests, monitoring the popularity of write requests is not required. Each virtual disk image of the running VM is divided into chunks of fixed size and the data organization agent records each chunk's access frequency. If the access frequency exceeds a predefined threshold for a particular chunk, the entire chunk will be outsourced to the replacement device. All the subsequent accesses to this chunk will be served by the replacement device, which removes their I/O involvement with the migration process. By submitting read-only requests, the migration module usually scans the entire virtual disk files. Most of these requests will only be issued once, except for requests that read dirty blocks of data (see Figure 1, Label 5).

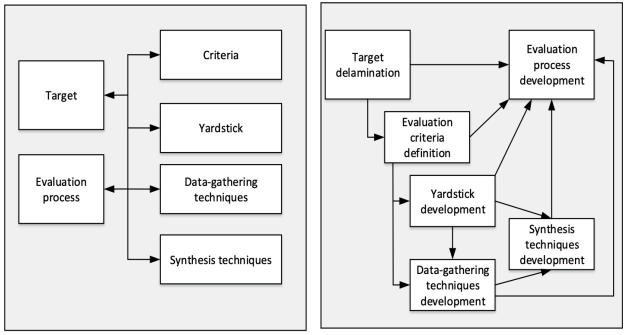

Figure 2. Components of an Evaluation and the Interrelationships between them [26].

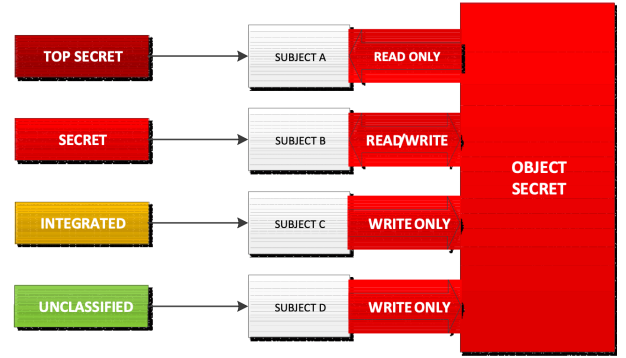

Figure 3. The Relationship Between Objects and Subjects.

This paper focuses on adopting evaluation theory to define the research system machine and consequently identify a way to apply the integrity model in the design research framework, as discussed in the next section.

## 4. EVALUATION OF THE SYSTEM ARCHITECTURE: STATE MACHINE

One of the primary aims of the proposed integrity framework is to consider the full cloud integrity environment and to capture all potential integrity attributes and elements as evidence, including functional and non-functional elements. Evaluation is a key analytical process for all intellectual disciplines and it is possible to apply different types of evaluation methods to provide knowledge regarding the complexity and ubiquity of the CSPs [25]. This paper aims to obtain a set of essential evaluation components. In particular, the evaluation of the system architecture method has been applied to review the secure establishment framework using the identification of these evaluation components and an analysis of their weaknesses and strengths. Evaluation theory [26] is considered a theoretical foundation for developing a secure live VM migration framework. Its processes are shown in Figure 2, which represents an overview of the components of evaluation and their interrelations, helping to establish a clear pathway for this study. Reaching a comprehensive and reliable integrity level in live VM migration processes is the main reason for using the evaluation theory. Further, this theory offers a clear, formal description of the evaluation concepts, as listed below:

- Target: Integrity between CSPs and cloud service users (CSUs).
- Criteria: Integrity elements of the CSPs and CSUs that are to be evaluated.
- Yardstick/standard: The ideal secure live VM migration framework measured against the current secure live VM migration framework.
- Data-gathering techniques: Critical or systematic literature review needed to obtain data to analyze each criterion.
- Synthesis techniques: Techniques used to access each criterion and therefore, to access the target, obtaining the result of the evaluation.
- Evaluation process: A series of tasks and activities that are used to perform the evaluation.

### 4.1 System Architecture State Machine

The proposed framework in this research is a state machine framework. It consists of subjects, objects, access attributes, access matrix, subject functions, and object functions. Access attributes are defined as follows: Read, Write, Read and Write, and Execute (depicted in Figure 3).

The proposed model state machine is as follows:

1) $t \in T$, where T is sorted Quaternion, each member of T is t
2) $T = (a, B, c, D)$, where,
3) $a \subseteq (S \times O \times A)$,
4) $B$ is an access matrix, where $B_{ij} \subseteq A$ signifies the access authority of $S_i$ to $o_i$,
5) $c \in C$ is the access class function, denoted as $c = (C_s, C_o)$,
6) $D$ signifies the existing hierarchy on the proposed framework,
7) $S$ is a set of Subjects,
8) $O$ is a set of Objects,
9) $A = [r, w, a, e]$ is the set of access attributes,
10) e$e$: R × T → I × T shows all the roles in the proposed framework, in which e is the system response and the next state, R is the requests set, and I is the arbitrary set of requests, which is [yes, no, error, question]. In this study, the question is important because if the response is equal to the question, it means that the current rule cannot deal with this request.
11) $\omega = [e_1, e_2, \dots, e_3]$, ω is the list exchange data between objects.

$W(\omega) \subseteq R \times I \times T \times T$

$(R_k, I_m, T^*, T) \in W(w)$

if $I_m \neq$ Question and exit a unique J, $1 \leq j \leq s$, it means that the current rule is valid, subject and object also are valid because the object verifies the vTPM of the other object (attestee) by request (challenge) for integrity checking. Consequently, the result is, $(I_m, t*) = e_i(R_k, t)$, which shows for all the requests in the t there is a unique response, which is valid. Where, $a \subseteq (S \times O \times A)$ where S is a set of Subjects, O is a set of Objects, and $A = [r, w, a, e]$ is the set of access attributes,

12) $c_s$ is the security level of the subject (includes the integrity level $c_1(S)$ and category level $c_4(S)$ ). Figure 3 shows the security level in the proposed framework and the relationships between the subjects and objects. $c_o$ signifies the security function of objects. Figures 3 show the relationship between the entire subjects, objects, security functions, and security level of the proposed framework.

13) The integrity of the vTPM is highest in the state machine and lowest in the user agent. Therefore, the integrity level is $c_1(TPM), c_2(TA), c_3(IDP), c_4(RP)$ and level $c_5(UA)$; this study should prove that each state of the proposed framework is secure. It has been assumed that each state is secure except for state three (Data Plane), as shown in Figure 1. Therefore, if state three is secure, all the states are secure.

14) $\Sigma(R, I, W, z_0) \subset X \times Y \times Z$

15) $(x, y, z) \in \Sigma(R, I, W, z_0)$, if $(z_t, y_t, z_t, z_t - 1) \in W$ for each $t \in T$, where $z_0$ is the initial state. Based on the above definition, $\Sigma(R, I, W, z_0)$ is secure in all states of the system; for example, $(z_0, z_1, ..., z_n)$ is a secure state.

16) CW model has several axioms (properties) that can be used to limit and restrict the

17) state transformation. If the arbitrary state of the system is secure, then the system is secure. In this study, the simple-security property (SSP) [27] is adopted. This property states that an object at one level of integrity is not permitted to read an object of lower integrity.

18) $t = (a, B, c, D)$

19) Satisfies SSP if,

20) For all s∈S, s∈S ⇒ [(o∈ a (s: r, w)) ⇒ $(c_s(s), > c_o(o)]$, i.e., $c_1(s) \geq c_2(o), c_3(s) \supseteq c_4(o)$.
$c_1(G) \geq c_2(vTPM), c_1(IEU) \geq c_2(RP)$.

Based on Figures 1, 3, and the SSP axiom, all the objects of the proposed framework use two primary concepts to ensure the security policy is enforced: well-informed transactions and separation of duties. The integrity axiom is "no read down" and "no write up", which means a subject at a specific classification level cannot read and write to data at a lower or higher classification respectively. Star property, Discretionary security, and Compatibility property are other models that can be used to limit and restrict the state transformation, and they will be used in future work.

## 5. CONCLUSIONS AND FUTURE WORK

The proposed framework, called Kororā, is designed based on five agents running on the Xen privileged dom0 and communicating solely with the hypervisor. The cloud scenario for this paper is a public cloud environment, which means the tenants have the most responsibility and control over their systems; therefore, the risks are higher. Consequently, as a response to the research problem, this paper has represented a design system architecture of a secure live VM migration. For further study, two more agents, called Go Agent and Libvirt Agent will be added to the Kororā in order to support the proposed framework being run in VMs and Xen hypervisor, respectively. A prototype will be developed to prove the effectiveness of the Kororā.